\documentclass[12pt]{article}
\usepackage{graphicx}

\newcommand\pubnumber{DPF2013-243}
\newcommand\pubdate{\today}

\def\bnl{Brookhaven National Laboratory \\ PO Box 5000 \\ Upton, NY 11973}

\def\Title#1{\begin{center} {\Large #1 } \end{center}}
\def\Author#1{\begin{center}{ \sc #1} \end{center}}
\def\Address#1{\begin{center}{ \it #1} \end{center}}

\newcommand\pubblock{\rightline{\begin{tabular}{l} \pubnumber\\
         \pubdate  \end{tabular}}}
\newenvironment{Abstract}{\begin{quotation}  }{\end{quotation}}
\newenvironment{Presented}{\begin{quotation} \begin{center} 
             PRESENTED AT\end{center}\bigskip 
      \begin{center}\begin{large}}{\end{large}\end{center} \end{quotation}}
\def\Acknowledgments{\bigskip  \bigskip \begin{center} \begin{large}
             \bf ACKNOWLEDGMENTS \end{large}\end{center}}

\def\beq{\begin{equation}}
\def\eeq#1{\label{#1}\end{equation}}
\def\eeqn{\end{equation}}

\def\beqa{\begin{eqnarray}}
\def\eeqa#1{\label{#1}\end{eqnarray}}
\def\eeqan{\end{eqnarray}}

\let\bar=\overbar

\def\Dslash{\not{\hbox{\kern-4pt $D$}}}
\def\dslash{\not{\hbox{\kern-2pt $\del$}}}

\def\msb{{\bar{\ssstyle M \kern -1pt S}}}

\newcommand{\nue}{\mbox{$\nu_e$}}
\newcommand{\numu}{\mbox{$\nu_{\mu}$}}
\newcommand{\nutau}{\mbox{$\nu_{\tau}$}}
\newcommand{\thetaonetwo}{\mbox{$\theta_{12}$}}
\newcommand{\thetatwothree}{\mbox{$\theta_{23}$}}
\newcommand{\thetaonethree}{\mbox{$\theta_{13}$}}
\newcommand{\mysth}{\mbox{sin$^2(2\theta_{13})$}}
\newcommand{\dcp}{\mbox{$\delta_{CP}$}}
\newcommand{\dmsq}{\mbox{$\Delta m^2_{21}$}}
\newcommand{\Dmsq}{\mbox{$\Delta m^2_{32}$}}
\newcommand{\Dmsqthreeone}{\mbox{$\Delta m^2_{31}$}}

\begin{document}
\begin{titlepage}
\pubblock

\vfill
\Title{Observation of electron antineutrino disappearance by the Daya Bay Reactor Neutrino Experiment}
\vfill
\Author{Elizabeth Worcester for the Daya Bay Collaboration}
\Address{\bnl}
\vfill
\begin{Abstract}
This presentation describes a measurement of the neutrino mixing parameter, 
$\mysth$, from the Daya Bay Reactor Neutrino Experiment. Disappearance of 
electron antineutrinos at a distance of $\sim$2 km from a set of six reactors,
where the reactor flux is constrained by near detectors, has been 
clearly observed. The result, based on the ratio of observed to expected 
rate of antineutrinos, using 139 days of data taken between December 24, 
2011 and May 11, 2012, is
$\mysth = 0.089 \pm 0.010 \mathrm{(stat.)} \pm 0.005 \mathrm{(syst.)}$. 
Improvements in sensitivity from inclusion of additional data, spectral 
analysis, and improved calibration are expected in the future.
\end{Abstract}
\vfill
\begin{Presented}
DPF 2013\\
The Meeting of the American Physical Society\\
Division of Particles and Fields\\
Santa Cruz, California, August 13--17, 2013\\
\end{Presented}
\vfill
\end{titlepage}
\def\thefootnote{\fnsymbol{footnote}}
\setcounter{footnote}{0}

\section{Introduction}
It is well-established experimentally that the flavor composition
of neutrinos change as they propagate\cite{mixing}. In the three-neutrino framework, the
three flavor states ($\nue,\numu,\nutau$) are superpositions of the three
mass states ($\nu_1,\nu_2,\nu_3$). The PMNS\cite{pmns1,pmns2}
matrix describes this mixing with three mixing angles 
($\thetaonetwo,\thetatwothree,\thetaonethree$), which have all been
measured experimentally, and a CP-violating phase,
$\dcp$, which is unknown. 
The differences between the mass states have also been measured, though
the true mass hierarchy, i.e., the sign of $\Dmsq$, is unknown.

The Daya Bay collaboration has measured the value of $\mysth$ by
observing the disappearance of electron antineutrinos produced at the
Daya Bay nuclear power complex in Guangdong, China \cite{db1,db2}. 
The survival probability for electron antineutrinos is:
\begin{equation}
P_{sur} \approx 1 - \mysth \mathrm{sin}^2(\frac{\Dmsqthreeone L}{4E_{\nu}}) - 
\mathrm{cos}^4(\thetaonethree)\mathrm{sin}^2(2\thetaonetwo) 
\mathrm{sin}^2(\frac{\dmsq L}{4E_{\nu}})
\end{equation}
The Daya Bay experiment makes
use of functionally-identical antineutrino detectors (ADs) placed
at several hundred meters from the reactors to constrain the reactor flux
and at several kilometers from the reactors to observe antineutrino
disappearance. This presentation reports the
Daya Bay measurement of $\mysth$ using data taken between December 24, 2011
and May 11, 2012, which is described in \cite{db2}.

\section{The Daya Bay Experiment}

\begin{figure}[htb]
\centering
\includegraphics[height=2in]{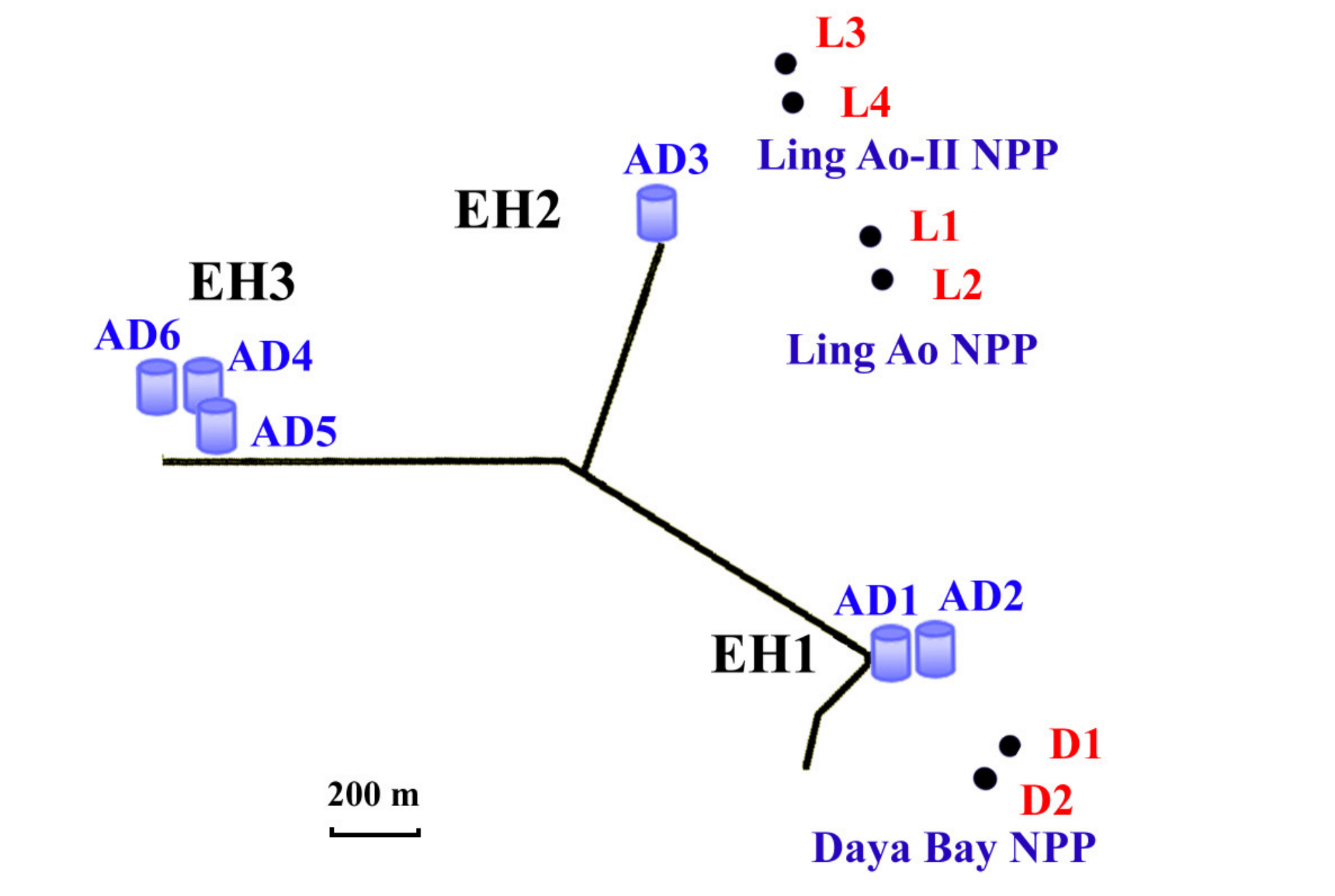}
\includegraphics[height=2in]{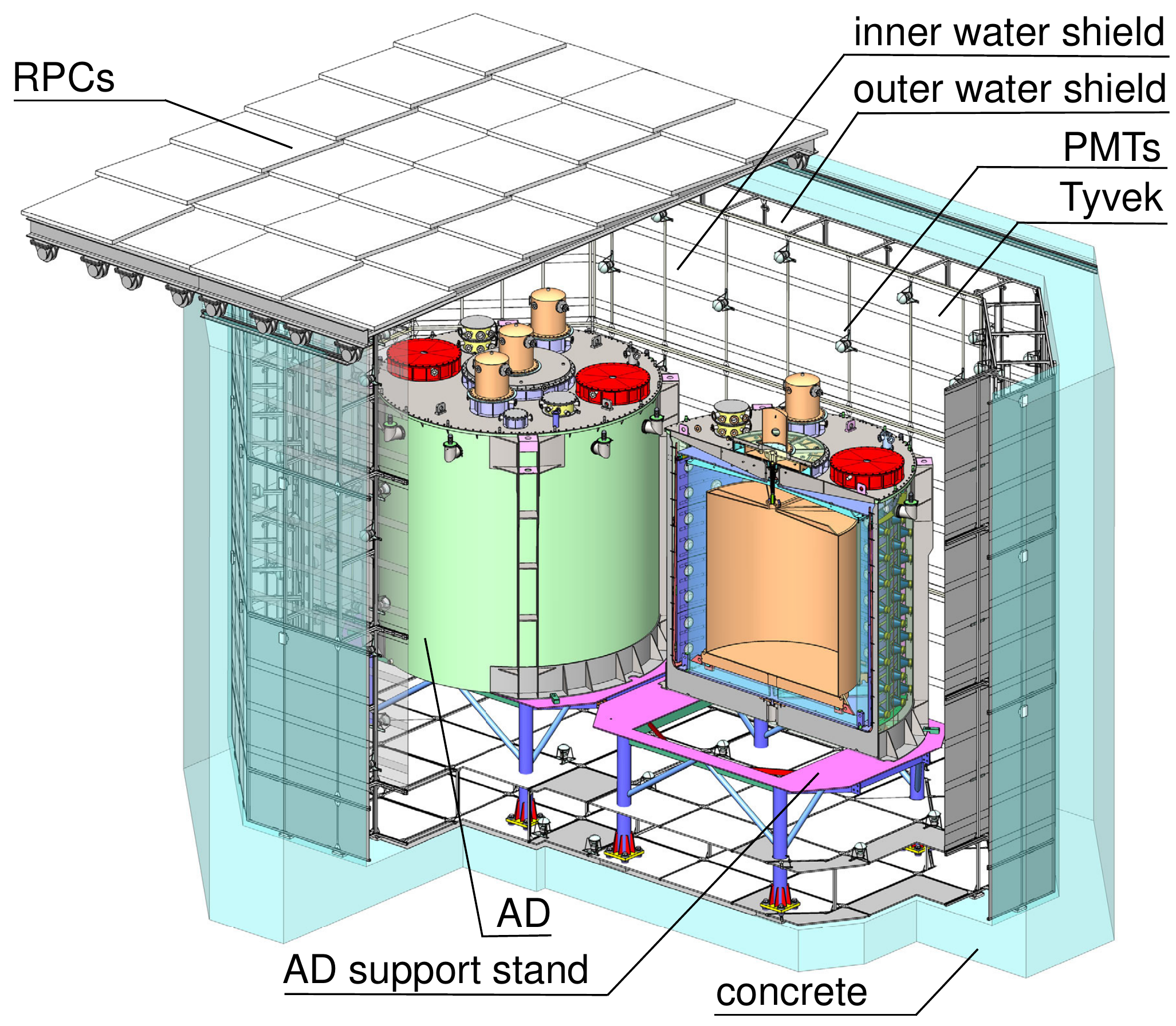}
\caption{(left) Schematic of the layout of Daya Bay experimental halls and
their location relative to the nuclear reactor cores. (right) Schematic
of AD location in water pool.}
\label{fig:dblayout}
\end{figure}

The Daya Bay nuclear power complex consists of six functionally-identical
pressurized water reactors, producing a total of up to 17.4 GW thermal power. The reactors
are grouped into three pairs, each of which is referred to as a ``nuclear
power plant'' (NPP): Daya Bay, Ling Ao, and Ling Ao II. The Daya Bay detectors
are located in underground experimental halls (EHs) that are
connected by horizontal tunnels and are located underneath nearby mountains,
which provide
shielding from cosmic-ray induced background. The experimental halls are
arranged such that EH1, containing two ADs, is near to Daya Bay NPP,
EH2, containing one AD during the period reported here, 
is near to Ling Ao and Ling Ao II NPPs,
and EH3, containing three ADs during the period reported here, 
is the far site. Figure \ref{fig:dblayout} (left panel) 
is a schematic of the experiment layout.

Daya Bay detects antineutrinos in the ADs using the inverse $\beta$-decay
(IBD) interaction in gadolinium-doped liquid scintillator\cite{gd1,gd2}.
In the IBD interaction, an antineutrino interacts with a proton to
produce a positron and a neutron. The positron produces a prompt 
scintillation signal
with energy related to that of the antineutrino. The neutron thermalizes
and is captured on gadolinium (Gd) with a characteristic capture time of 30 $\mu$s. This
prompt-delayed coincidence provides a distinct signature for antineutrino
interaction. 

The ADs are
three-zone detectors; the central region contains $\sim$20 tons of Gd-doped
liquid scintillator to detect IBD events, the middle region contains $\sim$20
tons of undoped liquid scintillator to detect $\gamma$ rays that escape the
target volume, and the outer region contains $\sim$40 tons of mineral oil to
shield the inner volumes from radioactive decay originating outside the
signal region.
Each AD is instrumented with 192
8-inch photomultiplier tubes (PMTs) installed on the inner wall of the outer
containment vessel and located in the mineral-oil region of the detector.
Each AD is equipped with three ``automated calibration units'' (ACUs), which 
allow deployment of calibration sources along three vertical axes of the 
detector. In each EH, the ADs are placed in a water pool which is 
instrumented with PMTs; the water pools act both as passive shielding
and as Cerenkov detectors to detect muons. Four-layer RPC modules are 
placed above each pool to provide additional muon information. Figure~\ref{fig:dblayout} 
(right panel) is a schematic of the placement of the ADs
inside the water pool in EH1.

The Daya Bay ADs are calibrated using sources deployed by the ACUs
in dedicated calibration periods:
a $^{68}$Ge source producing at-rest positrons, a
$^{241}\mathrm{Am}-^{13}\mathrm{C}$ source that produces 3.5 MeV neutrons,
a $^{60}\mathrm{Co}$ source that produces two $\gamma$s with total energy
$\sim$2.5 MeV, and an LED diffuser ball. The calibration also makes use
of spallation-neutron data taken simultaneously with IBD data during regular
physics data collection.

\section{Analysis}
Event selection consists of the removal of instrumental background
from the spontaneous emission of light by PMTs (``flashers''),
which is done by identifying the topology of this type of event,
removal of muon background, which is done by placing various requirements
on the energy deposit in the water pools and ADs, and selection of the
characteristic prompt-delayed IBD signature by requiring a prompt energy
deposit of 0.7-12 MeV, a delayed energy deposit of 6-12 MeV, and a capture
time in the range 1-200 $\mu$s. We further require no other signal $>0.7$ MeV
within $\pm$200~$\mu$s of the IBD candidate.
The result
presented here is based on six-AD data taken between December 24, 2011 and May 11, 2012.

We identify five sources of background to the IBD event sample. The largest
source of background is accidental coincidence, which contributes 1.5\%
of IBD candidates in the near halls and 4\% of IBD candidates in the far hall.
The other four sources of background, which are fast neutrons created by cosmic rays,
$\beta$-n decay of cosmogenic $^9$Li/$^8$He, capture on metal nuclei of
neutrons emitted by the $^{241}\mathrm{Am}-^{13}\mathrm{C}$ source in the ACUs, 
and $^{13}\mathrm{C(}\alpha,\mathrm{n})^{16}\mathrm{O}$
background, are all significantly smaller. The total background is 2\%(5\%)
in the near(far) ADs. 

Systematic uncertainties are quite small for this analysis because many uncertainties cancel in the
near-far ratio method employed by Daya Bay. The two largest uncorrelated, i.e., different
between ADs, systematic effects are from uncertainty in the delayed-energy selection
requirement and in the fraction of IBD neutrons that capture on Gd. The combined,
uncorrelated, detector-related uncertainty is about 0.2\%. The effect of uncorrelated uncertainty
in the reactor flux is 0.04\%.

\section{Results}
\begin{figure}[htb]
\centering
\includegraphics[height=2in]{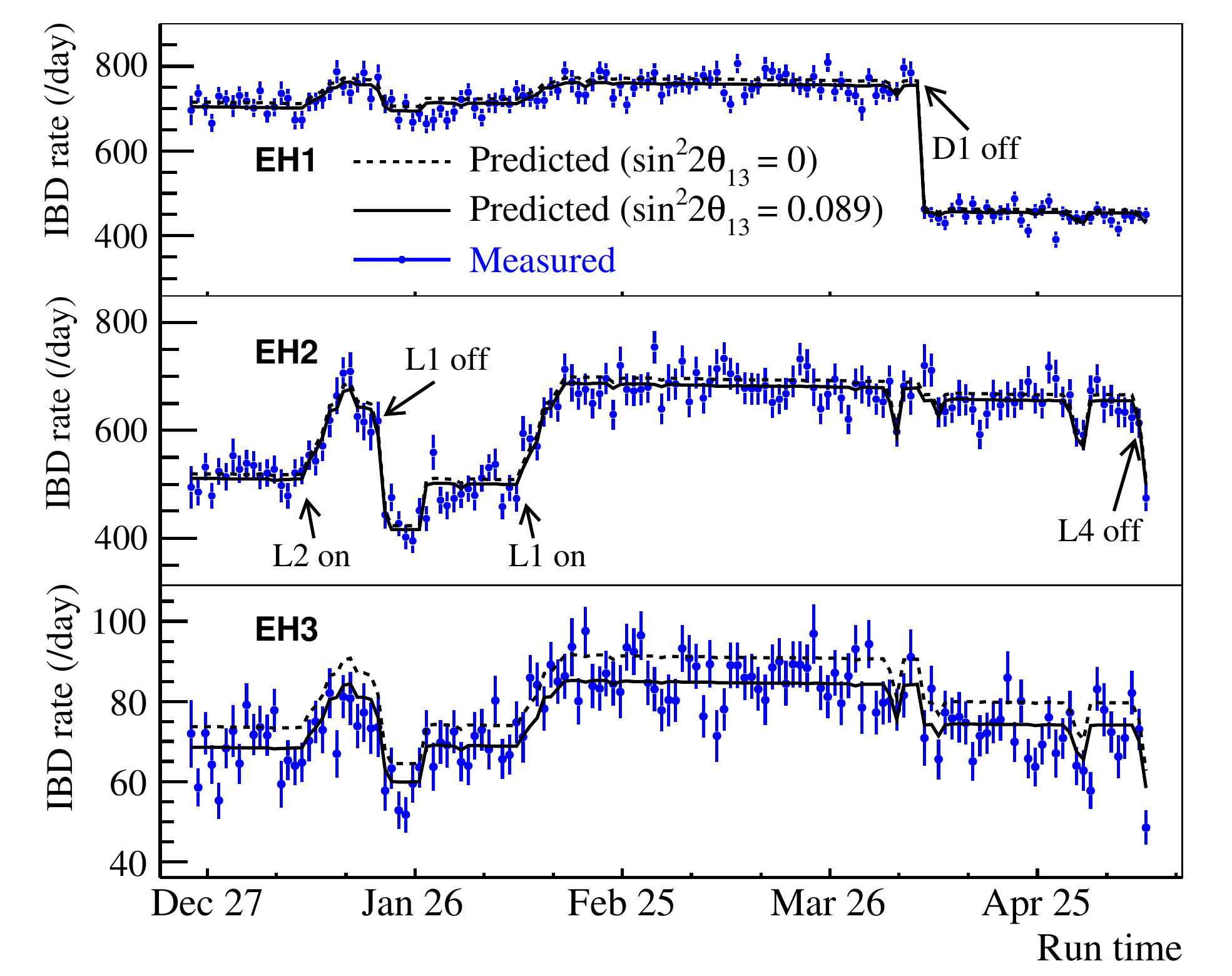}
\includegraphics[height=2in]{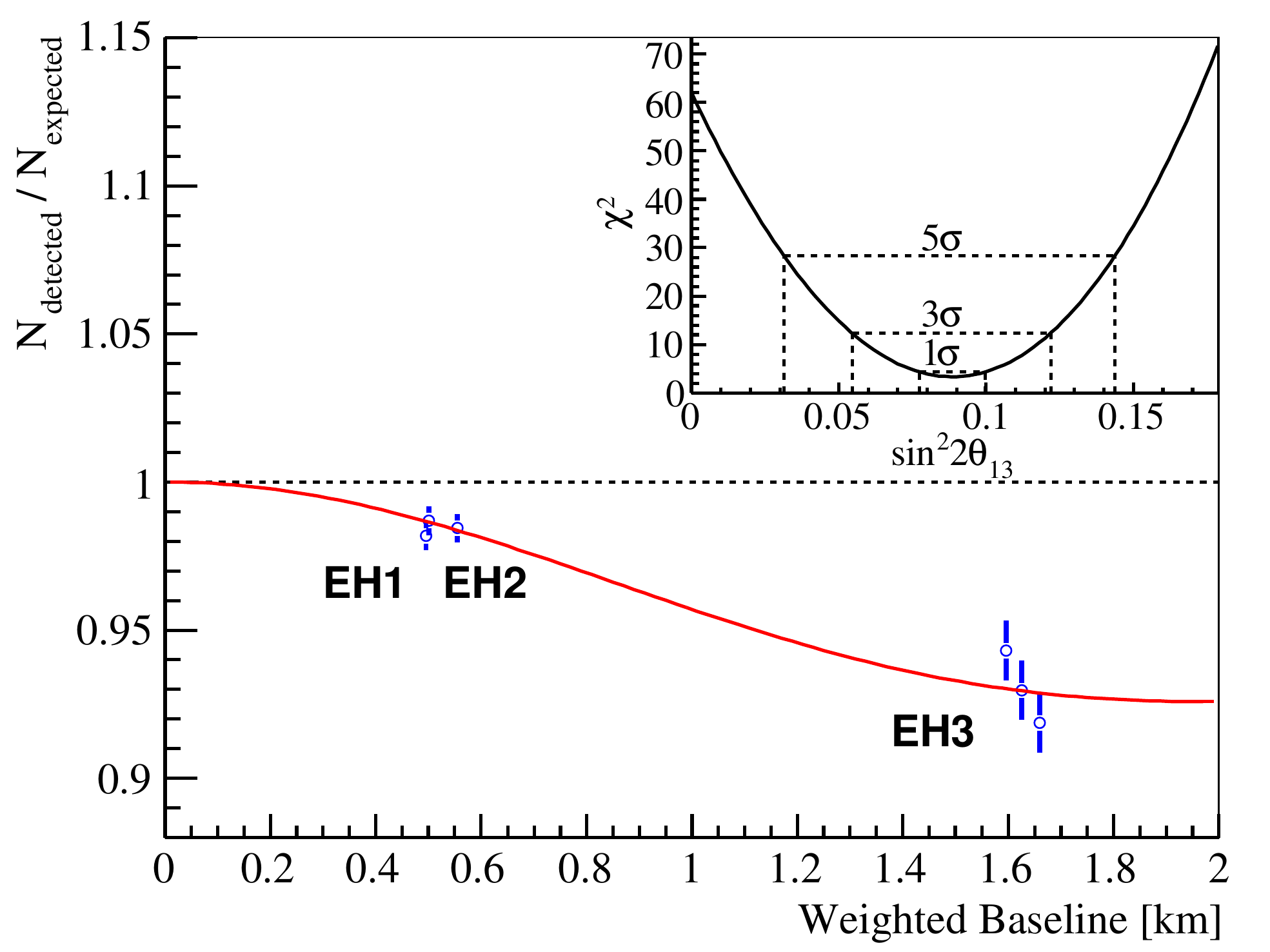}
\caption{(left) The measured daily average IBD rates per AD in the three experimental halls
as a function of time, compared to the expected reactor flux under the assumptions
described in the text. (right) Ratio of measured to expected antineutrino rate in each AD, 
assuming no oscillation.}
\label{fig:results}
\end{figure}

The left panel of Fig. \ref{fig:results} 
shows the detected antineutrino rate per AD as a function of time for
each experimental hall. For comparison, the predicted rates for a no-oscillation hypothesis
and for the best-fit value of $\mysth$ are shown, where the normalization of these curves
is determined by a fit to the data.
The fluctuations in rate are the result of various reactor cores
being powered on and off. The detected rate is strongly correlated with the reactor flux
expectations.

The value of $\mysth$ is determined using a standard $\chi^2$ approach, in which only the
rate of detected antineutrinos in each experimental hall is considered. Pull terms are used
to account for the correlation of the systematic errors. The quantity that is minimized is:
\begin{eqnarray}  \label{eqn:chi2}
 \chi^2 &=&
 \sum_{d=1}^{6}
 \frac{\left[M_d-T_d\left(1+  \varepsilon
 + \sum_r\omega_r^d\alpha_r
 + \varepsilon_d\right) +\eta_d\right]^2}
 {M_d+B_d}  \nonumber \\
 &+&
 \sum_r\frac{\alpha_r^2}{\sigma_r^2}
 + \sum_{d=1}^{6} \left(
 \frac{\varepsilon_d^2}{\sigma_d^2}
 + \frac{\eta_d^2}{\sigma_{B}^2}
 \right),
\end{eqnarray}
where $M_d$ are the measured IBD events of the $d$-th AD with its backgrounds subtracted, 
$B_d$ is the corresponding background, $T_d$ is a prediction of the antineutrino flux, 
$\omega_r^d$ is the fraction of IBD contribution of the $r$-th reactor to the $d$-th AD. 
The uncorrelated reactor uncertainty is $\sigma_r$ (0.8\%). The parameter 
$\sigma_d$ (0.2\%) is the uncorrelated detection uncertainty. The parameter $\sigma_{B}$ 
is the background uncertainty. The corresponding pull parameters are 
($\alpha_r, \varepsilon_d, \eta_d$). The absolute normalization, $\varepsilon$, is determined 
from the fit to the data. The best-fit result is
\begin{equation}
\mysth = 0.089 \pm 0.010 \mathrm{(stat.)} \pm 0.005 \mathrm{(syst.)},
\end{equation}
with a $\chi^2$/NDF of 3.4/4. This is the most precise measurement of $\mysth$ to date and
is in excellent agreement with other measurements from reactor- and accelerator-based
experiments\cite{dc,reno,minos,t2k}. Figure \ref{fig:results} (right panel) 
shows the ratio of measured to expected antineutrino
rate in each
AD, assuming no oscillation; the nearly 6\% deficit in the far experimental hall is clearly visible.

\section{Future Plans}
A spectral measurement of electron antineutrino oscillation amplitude and frequency at Daya Bay
was underway at the time of this presentation. The result has since been announced; the spectral
shape is consistent with the three-neutrino oscillation scenario, the value of $\mysth$ is
consistent with the result presented here, and the measurement of the mass-splitting, when
converted to a value of $\Dmsq$ assuming normal or inverted hierarchy, is consistent with the
result from atmospheric neutrinos. A paper describing these results is in preparation.

In fall 2012, a set of special calibration data was taken, 
including a 4$\pi$ calibration of one of the ADs
using the ``Manual Calibration System'' (MCS) \cite{mcs}. This data may be used to improve 
understanding of the energy response and absolute efficiency of the Daya Bay detectors. The
final two ADs were installed in fall 2012; Daya Bay has been taking data with all eight ADs
since October 2012.

\section{Summary}
The Daya Bay experiment has measured 
$\mysth = 0.089 \pm 0.010 \mathrm{(stat.)} \pm 0.005 \mathrm{(syst.)}$
by observing a deficit in the rate of electron antineutrinos at a distance $\sim$~2~km
from a nuclear reactor, relative to the rate expected with no oscillations. The is
the most precise measurement of $\mysth$ to date. Additional improvements in sensitivity
from inclusion of additional data, spectral analysis, and improved calibration are
expected.

\scriptsize
\Acknowledgments
The Daya Bay experiment is supported in part by the Ministry of Science and
Technology of China, the United States Department of Energy, the Chinese
Academy of Sciences, the National Natural Science Foundation of China, the
Guangdong provincial government, the Shenzhen municipal government, the China
Guangdong Nuclear Power Group, Shanghai Laboratory for Particle Physics and
Cosmology, the Research Grants Council of the Hong Kong Special Administrative
Region of China, University Development Fund of the University of Hong Kong,
the MOE program for Research of Excellence at National Taiwan University,
National Chiao-Tung University, NSC fund support from Taiwan, the U.S.
National Science Foundation, the Alfred P. Sloan Foundation, the Ministry
of Education, Youth and Sports of the Czech Republic, the Czech Science
Foundation, and the Joint Institute of Nuclear Research in Dubna, Russia.
We thank Yellow River Engineering Consulting Co., Ltd. and China railway
15th Bureau Group Co., Ltd. for building the underground laboratory. We are
grateful for the ongoing cooperation from the China Guangdong Nuclear Power
Group and China Light \& Power Company.

\end{document}